\documentclass[]{spie}

\pdfoutput=1 

\usepackage[]{graphicx,color,rotating}
\usepackage[]{amsmath}
\usepackage[]{amssymb}

\def\gtrsim{\mathrel{\hbox{\rlap{\hbox{\lower4pt\hbox{$\sim$}}}\hbox{$>$}}}}

\title{First Light with ALES: A 2-5 Micron Adaptive Optics Integral Field Spectrograph for the LBT}

\author{
Andrew J. Skemer\authorinfo{\supit{*} Contact: askemer@as.arizona.edu}\supit{a,b},
Philip Hinz\supit{a},
Manny Montoya\supit{a},
Michael F. Skrutskie\supit{c},
Jarron Leisenring\supit{a},
Oli Durney\supit{a},
Charles E. Woodward\supit{d}
John Wilson\supit{c}
Matt Nelson\supit{c}
Vanessa Bailey\supit{a},
Denis Defrere\supit{a},
and
Jordan Stone\supit{a}
\skiplinehalf
\supit{a} Steward Observatory, University of Arizona, Tucson, AZ, USA;
\skiplinehalf
\supit{b} University of California, Santa Cruz, Santa Cruz, CA, USA;
\skiplinehalf
\supit{c} University of Virginia, Charlotteville, VA, USA;
\skiplinehalf
\supit{d} Minnesota Institute for Astrophysics, University of Minnesota, Minneapolis, MN, USA;
}

\begin{document} 
\maketitle

\begin{abstract}
Integral field spectrographs are an important technology for exoplanet imaging, due to their ability to take spectra in a high-contrast environment, and improve planet detection sensitivity through spectral differential imaging.  ALES is the first integral field spectrograph capable of imaging exoplanets from 3-5 $\mu$m, and will extend our ability to characterize self-luminous exoplanets into a wavelength range where they peak in brightness.  ALES is installed inside LBTI/LMIRcam on the Large Binocular Telescope, taking advantage of existing AO systems, camera optics, and a HAWAII-2RG detector.  The new optics that comprise ALES are a Keplerian magnifier, a silicon lenslet array with diffraction suppressing pinholes, a direct vision prism, and calibration optics.  All of these components are installed in filter wheels making ALES a completely modular design.  ALES saw first light at the LBT in June 2015.
\end{abstract}

\section{INTRODUCTION\label{sec:intro}}
By spatially resolving exoplanets from their host stars, we can study their spectral energy distributions, and infer their bulk physical properties.  Typically, exoplanet imaging requires a large aperture, exquisite wavefront control from an adaptive optics system, and an infrared imaging camera.  Recently, integral field spectroscopy (IFS) has become a reliable alternative to conventional imagers\cite{2006NewAR..50..362L,2008SPIE.7015E..19H,2014PNAS..11112661M,2008SPIE.7014E..18B}.  IFS's enable spectroscopic characterization combined with the contrast enhancing power of 2-dimensional post-processing techniques, such as angular differential imaging\cite{2006ApJ...641..556M}.  Spectral information from IFS's can further improve planet detection sensitivity\cite{2014IAUS..299...48M}.

Until now, all planet imaging IFS's have been designed to operate in the near-infrared ($\sim$1-2$\mu$m), where optics and detectors are reasonably straightforward.  A competing technique to search for self-luminous exoplanets is to image in the mid-infrared (3-5$\mu$m), where gas-giant planets peak in brightness compared to their host stars (see Figure 1).  The Large Binocular Telescope (LBT), its low emissivity adaptive secondaries (LBTAO\cite{2011SPIE.8149E..02E}), the Large Binocular Telescope Interferometer (LBTI\cite{2014SPIE.9146E..0TH}) and its 1-5$\mu$m AO imaging camera (LMIRcam\cite{2010SPIE.7735E..3HS,2012SPIE.8446E..4FL}) have all been designed, in part, for exoplanet imaging.  This system already produces world-class planet sensitivities in the mid-infrared\cite{2012ApJ...753...14S,2014SPIE.9148E..0LS}.  The addition of a mid-infrared IFS will improve its planet imaging sensitivity, while also allowing spectroscopic characterization of exoplanets at a critical wavelength.

\begin{figure}[htbp]
  \hbox{
    \hspace{1.4in}
      \includegraphics[angle=0,width=0.7\linewidth]{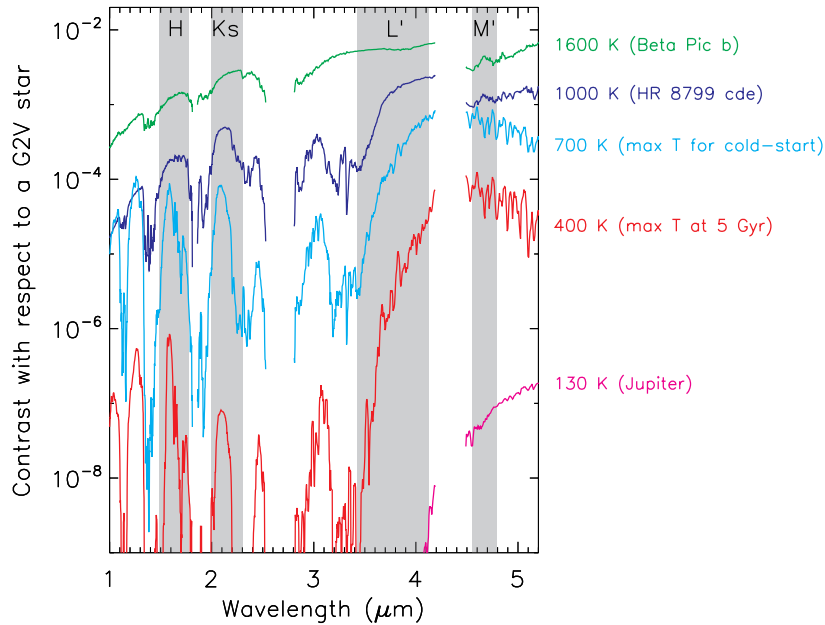}}
      \vspace{0.0in}
      \caption{Reproduced from Ref~[\citenum{2014ApJ...792...17S}]:  Characteristic examples of exoplanet-to-star contrasts (i.e. flux ratios) as a function of wavelength\cite{2011ApJ...737...34M,2003ApJ...596..587B}, showing (1) that gas-giant exoplanets can be detected with lower contrasts in the mid-infrared (3-5$\mu$m) than in the near-infrared (1-2$\mu$m), and (2) that this difference increases at lower temperatures.  While the planets that have been directly imaged to date ($\beta$ Pic b and HR 8799 c, d and e on this plot) are relatively warm (1600 K and 1000 K, respectively), it is likely that the majority of self-luminous exoplanets are much cooler.  Planets that formed by core-accretion (approximated by the ``cold-start'' models\cite{2008ApJ...683.1104F}) are never hotter than $\sim$700 K.  Planets around average-aged stars (5 Gyr) are never hotter than $\sim$400 K, regardless of formation history.  Jupiter, which may be a ubiquitous outcome of planet formation, is only $\sim$130 K. 
	  } \label{fig:L'just}
\end{figure}

In this paper, we describe the development of Arizona Lenslets for Exoplanet Spectroscopy (ALES), a fast-tracked IFS upgrade to the LBTI/LMIRcam adaptive optics imager.  ALES takes advantage of an existing AO system, dewar, camera optics, filter wheels, and HAWAII-2RG detector, along with 3 pairs of intermediate focal and pupil planes.  By integrating the IFS optics into mostly existing mechanisms, ALES is a modular design, capable of doing a wide variety of IFS science from $\sim$1.2-5$\mu$m.  The first mode of ALES, which is described here, has a 50x50 spaxel, $1.2\times1.2^{\prime\prime}$ field of view, with 2.8-4.2$\mu$m, R$\sim$20 spectroscopy.  Such a configuration is ideal for taking low-resolution spectra of exoplanets across broad molecular features.  It also covers the 3.1$\mu$m ice feature, PAH emission, and Br-$\alpha$.  Modest upgrades could allow modes at any wavelength from 1.2-5$\mu$m with spectral resolutions up to $\sim$1000, double the current spaxel field-of-view, and with spatial resolutions commensurate with the LBT's full 23-meter diameter resolving power.  In Section 2, we describe the design of ALES. In Section 3, we present first-light images.  In Section 4, we present a preliminary pipeline.  In Section 5, we tabulate basic properties.  In Section 6, we discuss future hardware upgrades.  We summarize in Section 7.

\section{Design\label{sec:design}}
\subsection{Integrating an IFS into LBTI/LMIRcam}
ALES was designed to fit into the existing LBTI/LMIRcam light-path (see Figure 2).  Two f/45 beams with an f/15 envelope enter LMIRcam, one for each side of the telescope.  Two biconic mirrors form two images of the pupil plane and two images of the focal plane.  In its normal configuration, LMIRcam's two focal planes (one intermediate and one at the detector) have a plate scale of $0.56^{\prime\prime}$/mm, corresponding to $0.01^{\prime\prime}$/pixel on the HAWAII-2RG detector.  Placing the lenslet array at the intermediate focal plane, each spaxel needs to produce a spectrum with a large number of pixels on the detector.  Since the intermediate and final focal planes have the same magnification, the spaxels will need to be physically much larger than the detector pixels.  As a result, achieving similar spatial sampling with the lenslet array as with the more finely pitched detector requires that the input beam be magnified.  The basic design for ALES is then a set of magnifying optics, a lenslet array, a disperser, and a corresponding bandpass filter that blocks the spectra from overlapping.

\begin{sidewaysfigure}[htbp]
  \vspace{-1.25in}
\centering
      \includegraphics[width=1.0\linewidth]{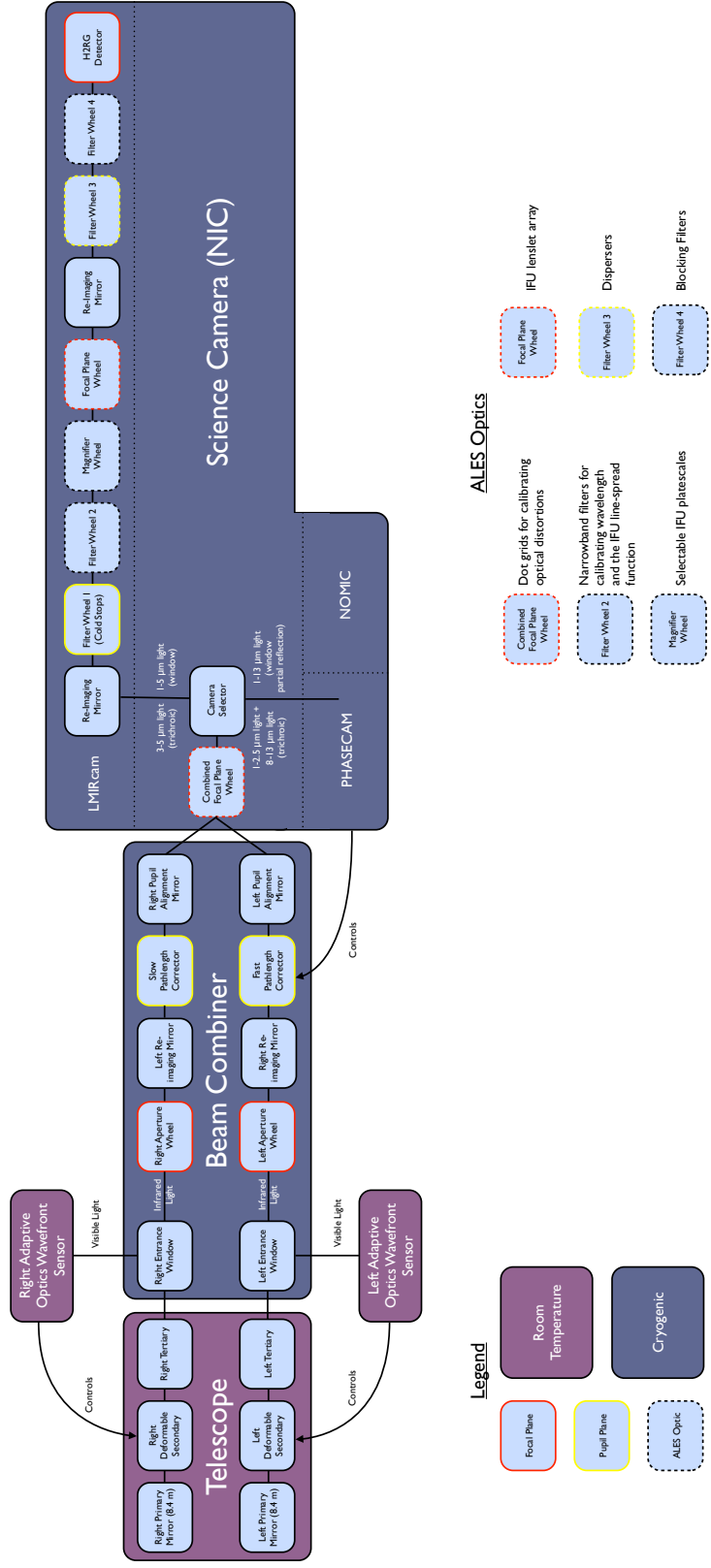}
      \vspace{-1.25in}
      \caption{Block diagram of LBTI / LMIRcam showing the full ALES light-path.  After reflecting off of three sets of warm telescope optics, light enters the cryogenic beam combiner and is directed into the cryogenic science camera, which contains LMIRcam.  The new optics that comprise ALES include a distortion grid, wavelength calibration filters, a Keplerian magnifier, a lenslet array, and a direct vision prism with a corresponding blocking filter.  All of the ALES optics are inserted by filter wheels, making the design completely modular.  Not shown: two flat mirrors and a ZnSe window between the beam combiner and the science camera.
	  } \label{fig:block}
\end{sidewaysfigure}

\subsection{Geometric Specifications}
In a lenslet-based IFS\cite{1995A&AS..113..347B}, the disperser is rotated with respect to the lenslet array so that the spectra are maximally separated.  For spaxels separated by $d$ pixels, the length of each spectrum is $l=d / sin(\theta)$ and the separation between spectra is $s=d\times sin(\theta)$, where $\theta=tan^{-1}(1/n)$ for a natural number, $n$.  An illustration of this concept is shown in Figure 3.  For ALES, we chose $d=20$ and $n=2$, giving $l=44.5$ pixels, $s=8.9$ pixels, and $\theta=26.57^{\circ}$.  This relatively large value of $s$ insures that the spectra will be easily separable, while $l$ is long enough to give reasonable spectral resolution for the exoplanet science case.  Leaving a similar buffer of $s=8.9$ pixels at the end of each spectra, our usable spectral range is 35.6 pixels.  LMIRcam has an existing blocking filter of 2.8-4.2$\mu$m that is suitable for exoplanet spectroscopy.  Therefore our disperser is designed to disperse 0.039$\mu$m/pixel.  Assuming a single wavelength spaxel PSF with a full-width at half maximum of $\sim$4-5 pixels (set by diffraction of the lenslets--see Section 2.3.2), this implies a spectral resolution of R$\sim$20, sufficient for characterizing the broad molecular features of exoplanets.

\begin{figure}[htbp]
\begin{center}
      \vspace{-0.25in}
      \includegraphics[angle=0,width=1.0\linewidth]{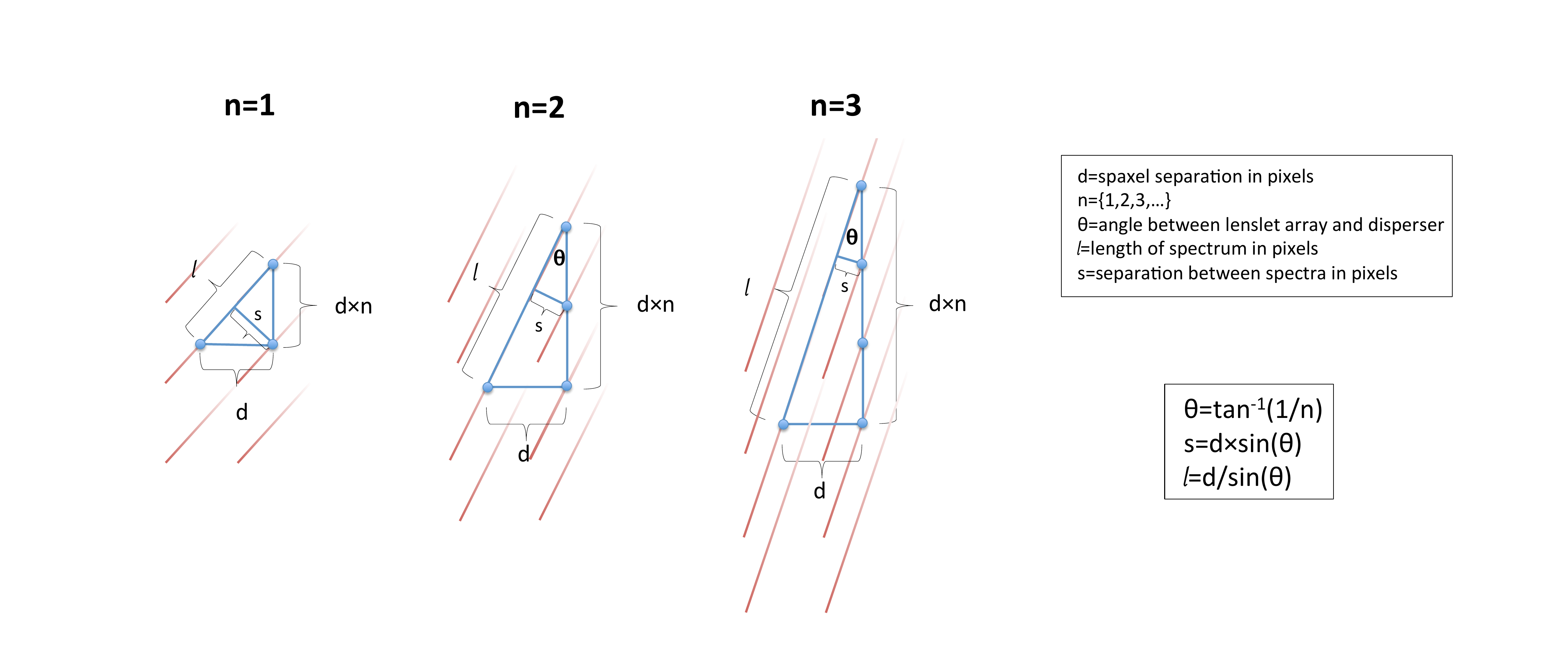}
\end{center}
      \vspace{0.0in}
      \caption{In lenslet based IFS's, the spectra (red) are rotated with respect to the spaxels (blue) in order to maximize the separation between adjacent spectra.  For a given spaxel separation, $d$ pixels, the length of each spectrum is $l=d / sin(\theta)$ and the separation between spectra is $s=d\times sin(\theta)$, where $\theta=tan^{-1}(1/n)$ for a natural number, $n$.
	  } \label{fig:rotation}
\end{figure}

The initial mode for ALES is 2.8-4.2$\mu$m imaging with an 8.4 meter diffraction-limited PSF, requiring $<$0.033''/spaxel for Nyquist sampling.  We chose 0.025''/spaxel, which allows Nyquist sampling to K-band.  Since each spaxel corresponds to $d=20$ pixels, and LMIRcam's nominal plate scale is 0.01''/pixel, we require a magnification of $\sim$8.

HAWAII-2RGs have 2048$\times$2048 pixels.  However, our current electronics only read out 1024$\times$1024 pixels.  An electronics upgrade is planned which will double our field-of-view (see Section 6).  For the 1024$\times$1024 pixel subarray with $d=20$ pixels/spaxel, we have room for 50$\times$50 spaxels, giving a field-of-view of 1.25''$\times$1.25''.

\subsection{Major Components}
\subsubsection{Magnifier}
\begin{figure}[htbp]
\begin{center}
      \includegraphics[angle=0,width=0.6\linewidth]{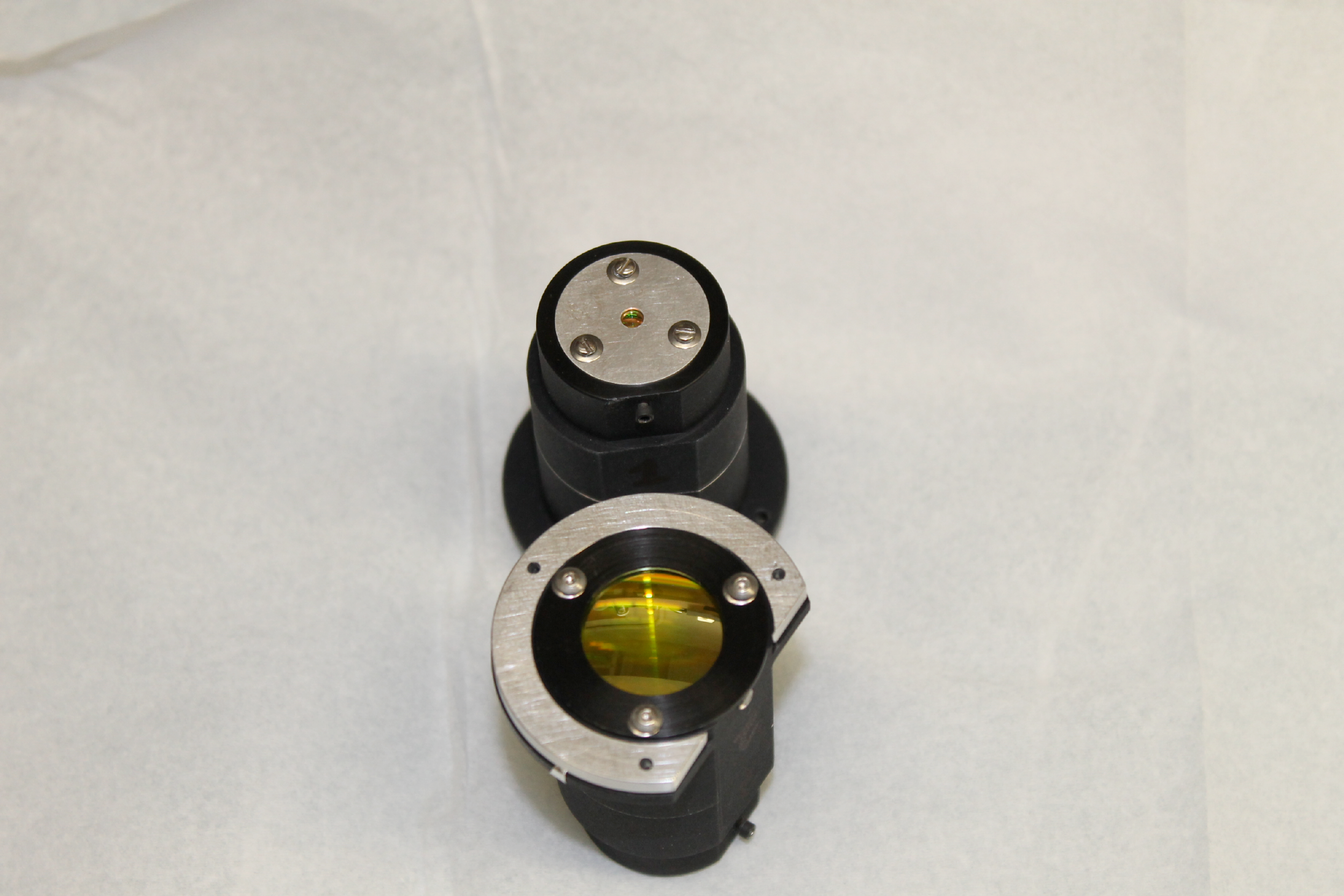}
\end{center}
      \vspace{-0.1in}
      \caption{Two identical 1-inch diameter lens tubes containing Keplerian magnifiers for ALES.  The bottom tube shows the entrance lens and the top tube shows the exit lens.
	  } \label{fig:magnifier}
\end{figure}

We installed a new cryogenic filter wheel between LMIRcam's first pupil plane and the intermediate focal plane to insert a tube of magnifying optics.  The wheel has 5 positions, one of which must remain open for conventional LMIRcam imaging.  The magnifier is a Keplerian telescope comprising two achromatic doublets.  Each doublet consists of ZnSe and CaFl with 2-5$\mu$m anti-reflection coatings.  The first three lenses have spherical surfaces and the last ZnSe lens is a diamond-turned asphere.  Each lens is mounted into the custom lens tube shown assembled in Figure \ref{fig:magnifier}.  Reference edges are used to align the optics, and a spring holds them in place as the assembly cools.

\subsubsection{Lenslet Array}
The heart of ALES is a custom lenslet array assembly, which re-images the focal plane into an array of equally spaced spots, which must have f/\#s no faster than f/15 to work with LMIRcam's downstream optics.  For many applications in the optical and near-infrared, lenslet array fabrication is relatively straightforward with photolithography.  Extending their operability to $\lambda\sim$5$\mu$m precludes the use of many commonly used optical materials.  Additionally, diffraction through the lenslet becomes more important at longer wavelengths.  Working with a manufacturer, \textit{Jenoptik Optical Systems}, we considered common 3-5$\mu$m materials, such as calcium fluoride, sapphire, diamond, and silicon.

In order to suppress diffractive cross-talk between lenslets, we opted to have pinholes aligned to the focus of each lenslet, as described in Ref.~[\citenum{2012SPIE.8446E..7UP}].  Normally, the lenslets can be on the front surface of a substrate, with chrome pinholes on the back surface.  However, our f/15 focal length constraint required an optic that was too thick to manufacture as a single substrate.  We opted to have two air-spaced substrates, with lenslets on the first substrate and pinholes on the second substrate.  Instead of having chrome deposited pinholes on a thin, flat, un-tilted substrate, which would create ghosts, we opted to etch through the second substrate.  From a manufacturing perspective, etching through-holes necessitated a silicon substrate.  The lenslet substrate has a 1.3-5$\mu$m anti-reflection coating on both sides.  The pinhole substrate has an \textit{Acktar} fractal black coating on both sides to block diffracted light and suppress scattered light.

\begin{figure}[htbp]
      \vspace{-0.3in}
\begin{center}
      \includegraphics[angle=0,width=0.95\linewidth]{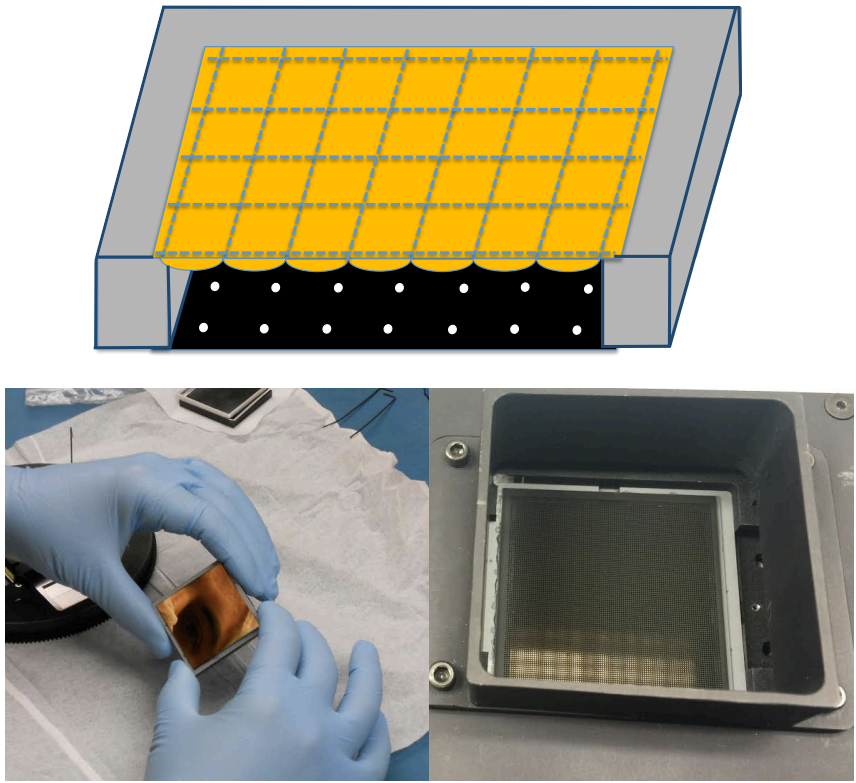}
\end{center}
      \vspace{-0.2in}
      \caption{Top: Cutout schematic of the lenslet array assembly (not to scale), which includes a lenslet array, spacer, and pinhole grid, all made from silicon.  Bottom Left: Frontside of the lenslet array (flat front with the lenslets interior) and the spacer during installation.  Bottom Right: Backside of the assembly (pinholes) as installed in the LMIRcam intermediate focal plane wheel.
	  } \label{fig:lenslet}
\end{figure}

The lenslet array and pinhole grid are aligned in the transverse direction and glued to a picture frame spacer, which separates the optics by a fixed distance.  All three pieces are made from silicon to avoid mechanical stresses during cooling.  Weep holes are cut into the sides of the picture frame to allow air to escape when the dewar is evacuated.  The picture frame is slightly oversized so that the mounting assembly in LBTI/LMIRcam is only in contact with the picture frame.  A schematic and photos of the assembled optic are shown in Figure \ref{fig:lenslet}.

\begin{figure}[htbp]
      \vspace{-0.05in}
\begin{center}
      \includegraphics[angle=0,width=0.75\linewidth]{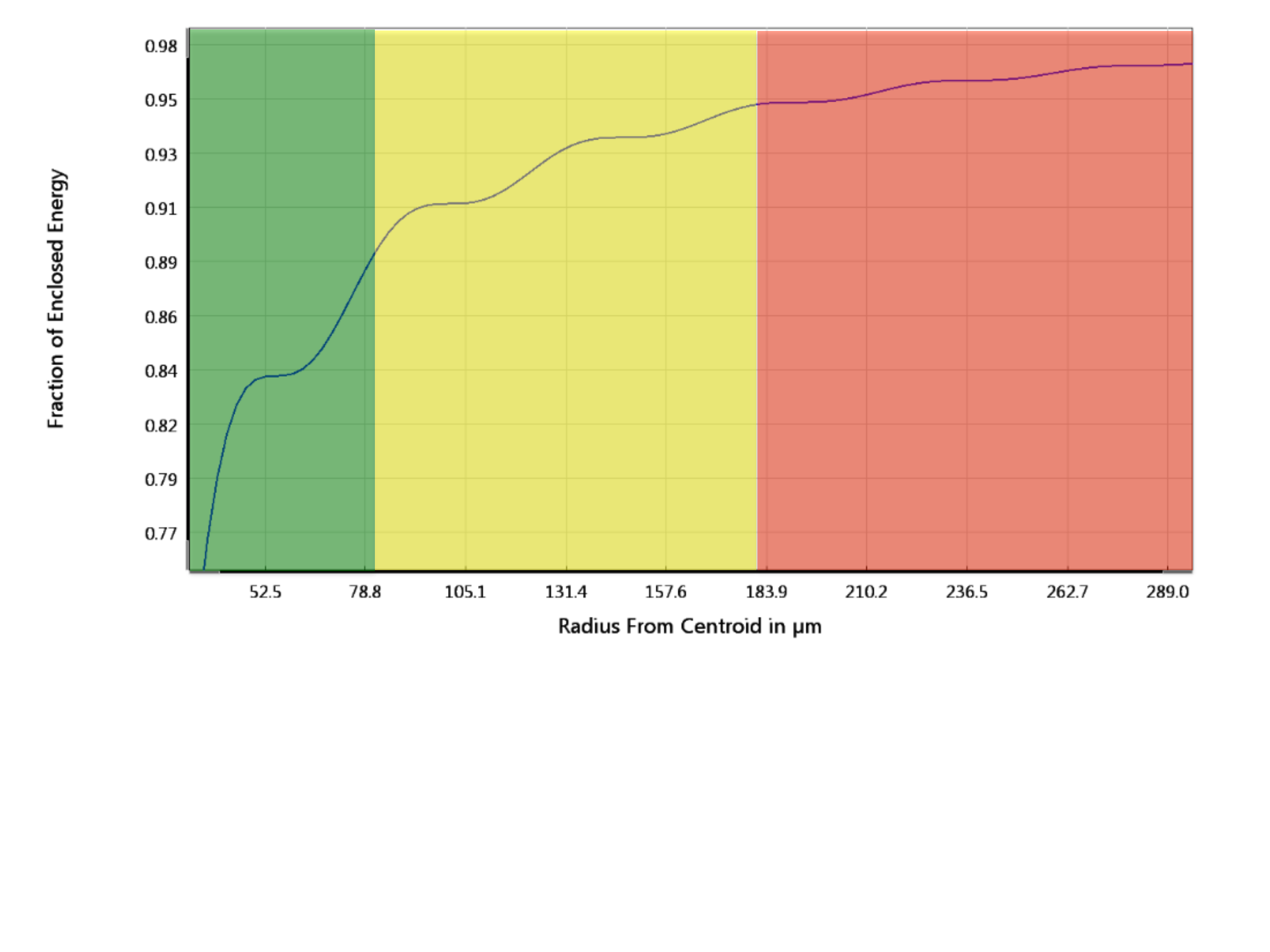}
\end{center}
      \vspace{-1.2in}
      \caption{Enclosed energy of a $\lambda=3\mu$m lenslet PSF at the location of the pinhole aperture.  In the green region, light is transmitted through the pinhole.  In the yellow region light is blocked by the pinhole, which would otherwise overlap with light from the adjacent dispersed spaxel.  In the red region, a very small fraction of light will go through adjacent pinholes ($<<$1\%); most of the circumference does not intersect the adjacent pinholes.
	  } \label{fig:enclosed energy}
\end{figure}

Each lenslet is a 360$\mu$m square (20 pixels on the HAWAII-2RG detector) on the backside of the lenslet array substrate.  The pinholes are designed to have the maximum diameter such that the rotated spectra do not overlap at the detector.  As described in Section 2.2, the spectra are separated by $s=8.9$ pixels, which corresponds to a pinhole size of 160$\mu$m.  Pinholes larger than this value will allow light from adjacent spaxels to overlap.  Pinholes smaller than this value will needlessly block usable light.  The theoretical pinhole throughput is 92\%, 89\%, 84\%, and 80\% at wavelengths of 2$\mu$m, 3$\mu$m, 4$\mu$m, and 5$\mu$m respectively. An enclosed energy curve is shown in Figure \ref{fig:enclosed energy}.

\subsubsection{Disperser and Blocker}
The disperser is a direct vision prism designed to disperse from 2.8-4.2 $\mu$m (matching an existing blocking filter in LMIRcam) across 35.6 detector pixels.  Our design uses a pair of ZnSe and sapphire prisms, which provide near linear dispersion across the desired band.  Each prism has a 2.8-4.2$\mu$m anti-reflection coating.  The prisms are mounted into a custom assembly which can be rotated precisely using a screw.  In principle, this can be used to optimize the angle between the lenslet array spaxels and the dispersion direction.  In practice, small movements of the prisms' filter wheel can be used to compensate misalignments.

Since LMIRcam was designed primarily as an imager, it does not have collimated light.  Therefore, we installed the prism assembly in a converging beam pupil plane filter wheel, which minimizes field dependent focal shifts from going through the varying thickness prisms.  The 2.8-4.2$\mu$m blocking filter is installed in a wheel immediately adjacent to the pupil plane wheel.

\subsection{Calibration}
\subsubsection{Narrowband Filters}
We purchased stock narrowband filters to calibrate the wavelength scale of the IFS.  Cryogenic scans of the filters give central wavelengths of 2.90$\mu$m, 3.36$\mu$m, 3.54$\mu$m, and 3.87$\mu$m.  The filters are narrow enough that they can be used to measure the single wavelength line-spread-function (the PSF of a single wavelength).  Our current batch of filters have inadequate blocking at the red end of our detector sensitivity ($\sim$4-5$\mu$m), where the night sky is very bright.  Thus, there is a spectrally dispersed diffuse background that affects measurements of the single wavelength spaxel PSF in the dispersion direction.

\subsubsection{Distortion Grid}
The magnifying optics are expected to impart a small distortion that is not present in LMIRcam's normal configuration.  At an intermediate focal plane that precedes LMIRcam, we installed a fixed-frequency dot grid that is magnified and imaged onto the lenslet array.  The dot grid is a stock part consisting of chrome dots spaced by 250$\mu$m on a 1.5 mm thick soda-lime glass substrate.  Soda-lime glass only has $\sim$30\% transmission from 3-4$\mu$m, but this is sufficient for imaging the dot grid.

\section{First Light\label{sec:first light}}
\begin{figure}[htbp]
\begin{center}
      \includegraphics[angle=0,width=1.0\linewidth]{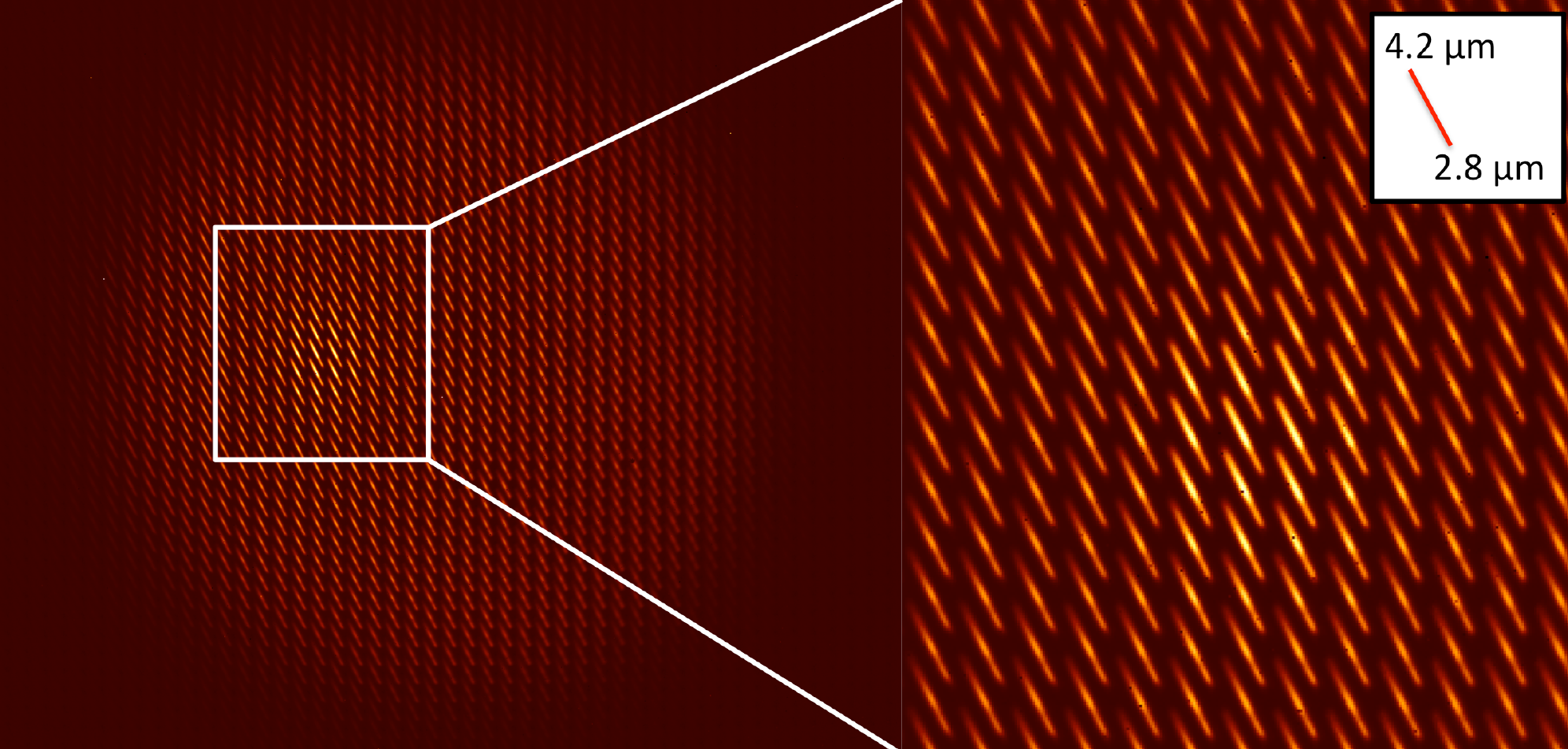}
\end{center}
      \caption{Left:Unprocessed ALES image of Io.  Right: Zoom-in on the Loki Patera volcano.
	  } \label{fig:Io}
\end{figure}

ALES was installed in LBTI's Nulling Infrared Camera (NIC) dewar in April-May 2015, during a two-month break in LBTI's observing schedule.  Other than a few quick cryogenic cycles to align the magnifying optics and lenslet array, very little testing was done in the lab.  NIC was reinstalled on the LBT at the beginning of June, and ALES had single-aperture first-light during a routine adaptive optics calibration period.  We observed a number of single and binary stars, as well as a few resolved sources to validate the system's basic properties.  An unprocessed IFS images of Io is shown in Figure \ref{fig:Io}.

\section{Pipeline\label{sec:pipeline}}
We have developed a simple pipeline that calibrates wavelength and extracts background subtracted data cubes, but we expect the software to evolve substantially.  Currently, our pipeline has the following basic steps:

\begin{enumerate}

\item Images of four narrowband filters, in series with the IFS, are used to identify spaxel locations as a function of wavelength.  The spaxel locations are fit by a 2D cubic polynomial to smooth spaxel centroiding errors.  The four known wavelength positions along with a theoretical dispersion curve for the direct vision prism, are used to determine an overall wavelength solution.  

\item Dark frames are subtracted from each raw data frame.  This step is particularly important because the spaxel spectra shift by up to one pixel with elevation changes.  Known bad pixels from a detector flat field are also removed.

\item On-source frames and sky background frames are registered to each other and then to a simulated image based on the wavelength solution.  

\item Each spectrum is extracted at 30 wavelengths perpendicular to the wavelength solution trace, with 6 pixel apertures.  The extracted spectra are reformatted into a data cube.

\item Sky background cubes are subtracted from on-source cubes.

\item Background subtracted cubes are divided by a master background cube that acts as an IFS flat-field.  

\item If the data were taken at high airmass ($\gtrsim$1.5), the individual slices may need to be shifted, particularly from $\sim$4.0-4.2$\mu$m, to compensate for mid-infrared atmospheric dispersion\cite{2008SPIE.7015E..5TK,2009PASP..121..897S}.  

\end{enumerate}

\begin{figure}[htbp]
\begin{center}
      \includegraphics[angle=0,width=1.0\linewidth]{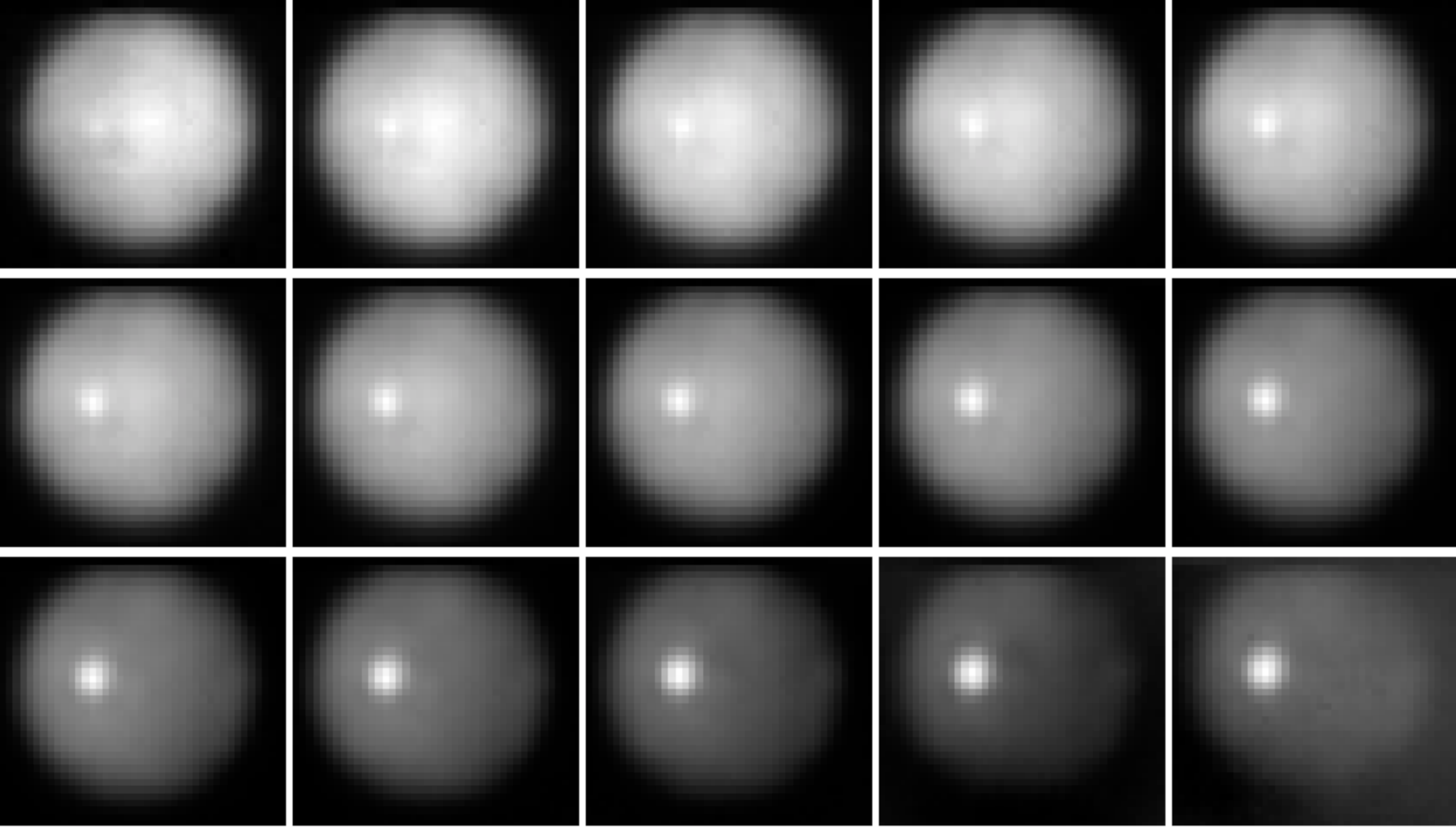}
\end{center}
      \caption{Extracted ALES images of Io with wavelengths ranging from 2.8$\mu$m (upper left) to 4.2$\mu$m (lower right) by steps of 0.1$\mu$m.  Light from Io's disk is primarily reflected Sun-light.  Thermal emission from the volcano peaks at longer wavelengths.
	  } \label{fig:Io extracted}
\end{figure}

Using our basic IFS pipeline, the ALES image of Io from Figure \ref{fig:Io} is extracted and presented as 15 wavelength calibrated images in Figure \ref{fig:Io extracted}.

\section{Measured Properties\label{sec:properties}}
\subsection{Throughput}
The magnifier and lenslet array have a combined 10 transmissive surfaces with a theoretical anti-reflection (AR) coating throughput of $\sim$92\%.  One of the calcium fluoride lenses cracked during installation and was replaced by a non-AR coated lens, which lowers theoretical throughput to $\sim$87\%.  The diffraction suppressing pinholes in the lenslet array assembly have a theoretical throughput of $\sim$85\% from 3-4$\mu$m (see Figure \ref{fig:enclosed energy}).  Therefore we expect the ALES throughput (magnifier and lenslet array) to be $\sim$74\%.

We imaged a laboratory source in LMIRcam's imaging mode and with the ALES optics (magnifier and lenslet array) to measure ALES's throughput. We find a throughput of $\sim$70\%, which is consistent with our theoretical estimate.  Note that the prisms and bandpass filter were not included in this analysis.  The prisms have a theoretical throughput of $\sim$99\%, and the bandpass filter has a measured throughput of $\sim$90-95\%, similar to imaging bandpass filters.

\subsection{Spectral Range / Resolution}
We compared an ALES spectrum of an F0 star with a telluric absorption template.  The spectrum is consistent with spectral resolutions of $\sim$15-20, covering the full L-band window from $\sim$2.9-4.1$\mu$m.  Spectral resolution is ultimately set by the diffraction-limited spaxel PSF size, which is significantly larger than two pixels, especially at longer wavelengths (see Figure \ref{fig:enclosed energy}).  ALES images through the narrowband filters described in Section 2.4.1 give single wavelength full-width at half maxima of $\sim$3, $\sim$3.5, $\sim$4 and $\sim$4 pixels at 2.90$\mu$m, 3.36$\mu$m, 3.54$\mu$m, and 3.87$\mu$m respectively.

\subsection{Plate Scale}
We observed the $\sim$0.3'' binary star, HD 213530 with LMIRcam in its imaging configuration, and with ALES.  LMIRcam has a well-characterized plate-scale\cite{2015A&A...576A.133M}, which we used to estimate ALES's plate-scale.  From 2.8-4.2$\mu$m, we find ALES has a plate-scale of 0.0261 ''/spaxel, with a standard deviation of 0.0001 ''/spaxel between wavelengths.  Geometric distortions (see Section 2.4.2) were not considered in this analysis.

\subsection{Image Quality}
\begin{figure}[htbp]
\begin{center}
      \includegraphics[angle=0,width=1.0\linewidth]{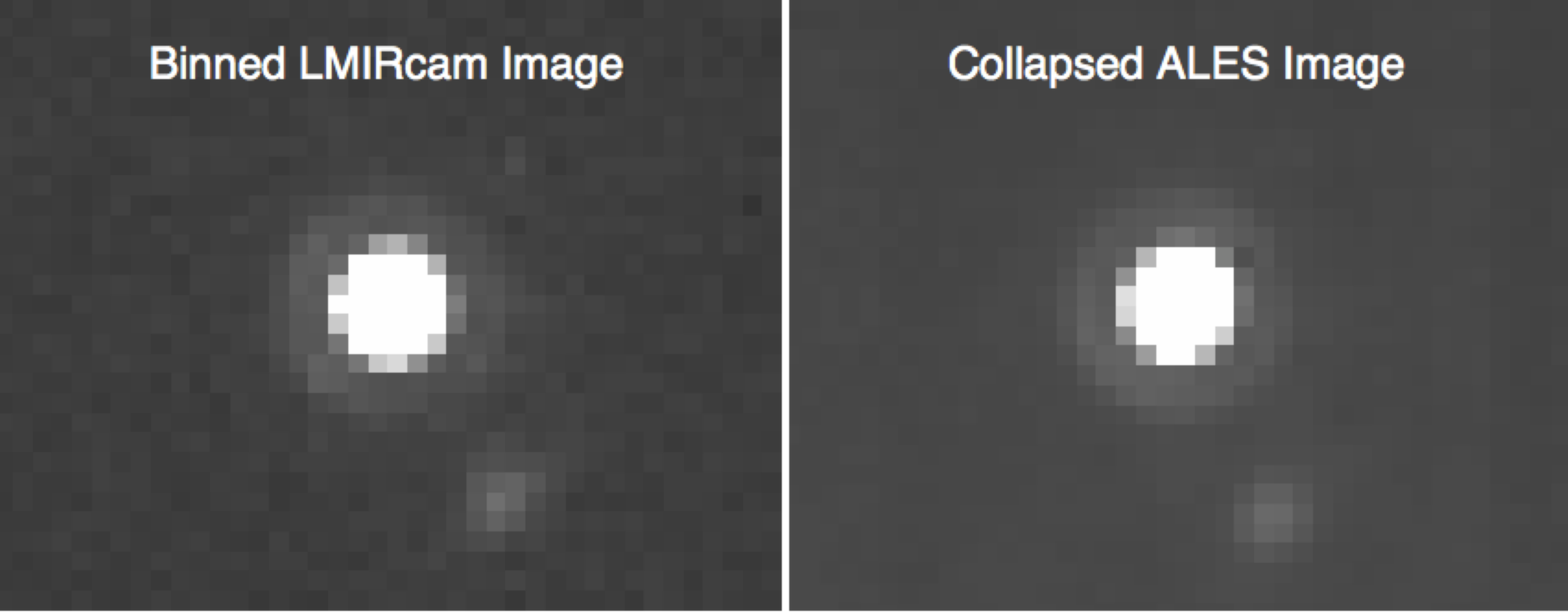}
\end{center}
      \caption{Images of a laboratory source (and associated ghost) taken with LMIRcam (left) and with ALES (right).  The LMIRcam image has been binned and interpolated to the ALES lenslet plate scale.  The ALES image is a collapsed ALES cube.  The bandpass for both images is 2.8-4.2$\mu$m.  ALES preserves LMIRcam's excellent image quality.
	  } \label{fig:image quality}
\end{figure}

We imaged a laboratory source with LMIRcam and ALES to validate ALES's diffraction-limited performance.  In Figure \ref{fig:image quality}, we show both images, which are qualitatively similar.  Quantitatively, defining the LMIRcam image as 100\% Strehl ratio, we measure a 105\% Strehl ratio for ALES (the excess is probably due to interpolation, or a slight focus offset between LMIRcam and ALES).  This result demonstrates that the ALES optics preserve LMIRcam's excellent image quality.

\subsection{Crosstalk}
The Io image in Figure \ref{fig:Io}, which is background subtracted but otherwise unprocessed, shows peak counts of $\sim$600-1000 digital units from 3-4$\mu$m in the center of the frame.  Pixels between adjacent spectra have an average of 8 counts.  Assuming the line-spread function continues to decline as it intersects an adjacent spaxel, this implies an average cross-talk of $<$1\%, which validates the use of the pinhole grid behind the lenslet array.

Near the edges of the field, astigmatism degrades the spaxel spots, and cross-talk becomes non-negligible.  The astigmatism is due to a biconic mirror that re-images the lenslet spots onto the detector.  In LMIRcam's normal imaging configuration, two re-imaging biconic mirrors each have astigmatism which are designed to cancel.  The astigmatism near the edges of the field could be suppressed by using longer focal length lenslets, which would illuminate less of the second biconic.  This is discussed in Section 6 as a possible future upgrade.

\subsection{Sensitivity}
We observed an L=6 magnitude A5 star with ALES.  After extracting the spectra into a data cube, we collapsed the cube into a single image with a bandpass of 2.9-4.2$\mu$m.  We fit the star image with a 2-dimensional Gaussian, and scaled the detection significance to 5$\sigma$ in 1 hour (30 minutes on-source, 30 minutes of background).  We find a sensitivity of L=16.4 mags.  The data were acquired during an AO engineering night, where the PSF had a lower Strehl ratio than normal, and LBTI's cryogenic beam combiner was warm, which contributes thermal background.  The normal instrument configuration, and the addition of dual-aperture ALES observations, will significantly improve sensitivity. 

\section{Future Hardware Improvements\label{sec:improvements}}
Because ALES has a modular design, additional IFS modes are relatively straightforward to install.  The lenslet array has an anti-reflection coating covering the H, K, L and M atmospheric windows, and its wheel has room for an additional lenslet array with, for example, a different spaxel size.  The magnifying wheel has room for four magnifying tubes.  We plan to replace the refractive optics with a reflective design\footnote{This is possible without vignetting from a central obscuration because LBTI's pupil is a scaled LBT pupil, comprising two off-center primaries}.  Magnifications of 2, 4, 8, and 16 will allow a variety of fields-of-view and spaxel plate scales, including an interferometric plate-scale for the L and M bands.  Additional low-spectral resolution modes require matched direct vision prisms and blocking filters in existing wheels.  Higher spectral resolution modes are possible using grisms, similar to LMIRcam's long-slit grisms\cite{2012SPIE.8450E..3PK}.

Currently, LMIRcam reads only 1024x1024 pixels of its full 2048x2048 detector array.  An upgrade of LMIRcam's electronics to the Teledyne SIDECAR/SAM is currently in progress, which will double ALES's field of view.  Near the edges of the field, we expect astigmatism will reduce spectral resolution and increase inter-spaxel crosstalk.  This can be mitigated with a longer focal length lenslet array.

Two sets of coronagraphs are installed in series with ALES, but have not been tested with ALES yet.  A pair of vortex coronagraphs\cite{2014SPIE.9148E..3XD} is installed in the ``Combined Focal Plane Wheel'' (see Figure 2) and a pair of vector apodized phase plates\cite{2014IAUS..299...40K} also serve as a cold stop in ``Filter Wheel 1'' (see Figure 2).

\section{Summary\label{sec:Summary}}
Integral field spectrographs are now frequently used to obtain spectra of directly imaged exoplanets.  However, the complexity of thermal infrared adaptive optics, thermal infrared cameras, and lenslet diffraction have precluded an instrument that operates from 3-5$\mu$m, where self-luminous gas-giants peak in brightness.  We report the concept, design, and first-light properties of ALES, a fast-tracked, 2-5$\mu$m IFS upgrade to LBTI/LMIRcam.  ALES's initial mode is a 2.8-4.2$\mu$m, R$\sim$20 IFS for characterizing exoplanets.  Our first light results demonstrate that ALES is performing as expected and is ready for high-contrast imaging of exoplanets.

\section*{ACKNOWLEDGMENTS}
The authors thank Marshall Perrin and Bruce Macintosh for useful discussions about IFS design.  A.S. is supported by the National Aeronautics and Space Administration through Hubble Fellowship grant HSTHF2-51349 awarded by the Space Telescope Science Institute, which is operated by the Association of Universities for Research in Astronomy, Inc., for NASA, under contract NAS 5-26555.  ALES hardware is funded by the University of Arizona Technology Research Initiative Fund, the Hubble Fellowship program, and the University of Minnesota.  The LBT is an international collaboration among institutions in the United States, Italy and Germany. LBT Corporation partners are: The University of Arizona on behalf of the Arizona university system; Istituto Nazionale di Astrophisica, Italy; LBT Beteiligungsgesellschaft, Germany, representing the Max-Planck Society, the Astrophysical Institute Potsdam, and Heidelberg University; The Ohio State University, and The Research Corporation, on behalf of The University of Notre Dame, University of Minnesota and University of Virginia.  The Large Binocular Telescope Interferometer is funded by the National Aeronautics and Space Administration as part of its Exoplanet Exploration program.  LMIRcam is funded by the National Science Foundation through grant NSF AST-0705296.

\bibliography{database}
\bibliographystyle{spiebib}

\end{document}